\documentclass[prc,aps,nofootinbib,showkeys,showpacs,twocolumn]{revtex4} 

\usepackage{epsfig}
\usepackage{graphicx}
\begin{document}
\date{\today}

\title{Exact stochastic simulation of dissipation and non-Markovian effects in open quantum systems}

\author{Denis Lacroix.} 
\affiliation{GANIL, CEA et IN2P3, BP 5027, 14076 Caen Cedex, France}
\begin{abstract}
The exact dynamics of a system coupled to an environment can be described 
by an integro-differential stochastic equation of its reduced density.
The  influence of the environment is incorporated through a mean-field which is both stochastic and non-local in time 
and where the standard two-times correlation functions of the environment appear naturally.
Since no approximation is made, the presented theory incorporates exactly dissipative 
and non-Markovian effects. Applications to the spin-boson model coupled to a heat-bath with various coupling regimes 
and temperature show that the presented stochastic 
theory can be a valuable tool to simulate exactly the dynamics of open quantum systems. 
Links with stochastic Schroedinger equation method and possible extensions to "imaginary time"
propagation are discussed. 
\end{abstract}


\maketitle

\section{Introduction}

Numerous concepts in our understanding of quantum mechanics have emerged from 
the understanding and description of a system coupled to an
environment: measurement, decoherence, appearance of classical world, irreversible process and 
dissipation... All these phenomena which are often encompassed
in the "theory of open quantum systems", bridge different fields of physics and 
chemistry \cite{Wei99,Joo03,Bre02}.   
During the past decade, important advances have been made in the approximate and 
exact description of system embedded in an environment using stochastic methods. 
Recently the description of open quantum systems by Stochastic Schroedinger Equation (SSE) 
has received much attention 
\cite{Ple98,Bre02,Sto02}. Nowadays, Monte-Carlo wave-function techniques are
extensively used to treat master equations  in the weak coupling and/or 
Markovian limit\cite{Dal92,Dum92,Gis92,Car93,Cas96,Rig96,Ple98,Gar00,Bre02}.

Large theoretical efforts are actually devoted to the introduction of
non-Markovian effects. Among the most recent approaches, one could mention
either deterministic approaches like Projection Operator techniques \cite{Bre99,Bre01}
or stochastic methods like   
Quantum State Diffusion (QSD) \cite{Dio96,Dio98,Str99,Str05} where   
non-Markovian effects are accounted for
through non-local memory kernels and state vectors evolve
according to integro-differential stochastic equations. In some cases, these
methods are shown to be exact \cite{Sto02,Sha04}.

Recently, alternative exact methods\cite{Bre04c,Lac05} have been developed to treat the system+environment 
problem that avoid evaluation of non-local memory 
kernel although non-Markovian effects are accounted 
for exactly. 
However, up to now, only few applications of these exact techniques exist
\cite{Bre04a,Bre04b,Bre04c,Lac05,Zho05,Bre07}. In all cases, 
accurate description of the short time dynamics is obtained but long time evolution can 
hardly be described  
due to the large increase of statistical errors with time.
Although, the possibility to simulate exactly the dissipative
dynamics of open quantum systems is already an important step, the challenge to describe long-time 
evolution is highly desirable to make the techniques more powerful and versatile.

\section{Exact stochastic equation for the reduced system density}

\subsection{Introduction}

In the present work, starting from the exact stochastic formulation of ref. \cite{Lac05} and projecting out the 
effect of the environment,  an equation of motion for the reduced system dynamics is obtained 
where the environment effect is incorporated through a stochastic mean-field which turns out 
to be non-local in time. Advantages of the new stochastic theory for the description of  long-time 
evolution of open quantum systems are underlined.
We consider here a system (S) + environment (E) 
described by a Hamiltonian
\begin{eqnarray}
H = h_S + h_E + h_I, 
\end{eqnarray}   
where $h_S$ and $h_E$ denote the system and environment Hamiltonians respectively 
while $h_I$ is responsible for the coupling. Here we assume that the interaction writes 
\begin{eqnarray}
h_I =  \mathbf{Q} \otimes \mathbf{B},
\end{eqnarray} 
where $\mathbf{Q}\equiv f(\{Q_i\}_{i=1,n_S})$ 
and $\mathbf{B}\equiv g(\{B_i\}_{i=1,n_B})$ correspond to functions of two  
sets of operators of the system and environment, respectively. In particular, this definition 
includes non-linear couplings.
For the sake of simplicity, we assume 
an initial separable density $D(t_0) = \rho_S(t_0) \otimes \rho_B(t_0)$. As will be discussed below 
this assumption could be relaxed. 
The exact 
evolution of the system is described by the Liouville von-Neumann equation $i\hbar \dot D = \left[H,D\right].$
Due to the coupling, the simple separable 
structure of the initial condition is not preserved in time. It has however been realized 
recently in several works using either SSE or path integral 
techniques  that the exact density of the total system $D(t)$ could be obtained as an average over
simple separable densities, i.e. $D(t) = \overline{\rho_S(t) \otimes \rho_B(t)}$. {
In its simplest version, the stochastic process takes the form  \cite{Lac05}
\begin{eqnarray}
\left\{
\begin{array} {lll}
d\rho_S &=& \frac{dt}{i\hbar}[h_S ,\rho_S] +  
d\xi_S \mathbf{Q} \rho_S + d\lambda_S 
\rho_S \mathbf{Q}
\\
\\
d\rho_E &=& \frac{dt}{i\hbar}[h_E,\rho_E] +  
d\xi_E \mathbf{B}\rho_E + d\lambda_E \rho_E \mathbf{B}
\end{array}
\right.
\label{eq:stocmfsimple}
\end{eqnarray}
where the Ito convention for stochastic calculations is used\cite{Gar85}. 
$d\xi_{S/E}$ and $d\lambda_{S/E}$ denote  Markovian Gaussian stochastic variables with zero means 
and variances
\begin{eqnarray}
\overline{d\xi_S d\xi_E} &=& \frac{dt}{i\hbar},~~~~\overline{d\lambda_S d\lambda_E} = -\frac{dt}{i\hbar}, \label{eq:noise1} \\
\overline{d\xi_S d\lambda_E} &=& \overline{d\lambda_S d\xi_E} = 0.
\label{eq:noise2}
\end{eqnarray} 
The average over stochastic paths described by Eqs. (\ref{eq:stocmfsimple}) match the exact evolution. 
Indeed, assuming that at time $t$ the total density writes $D(t) = \rho_S(t) \otimes \rho_B(t)$, the average 
evolution over a small time step $dt$ is given by
\begin{eqnarray}
\overline{dD} = \overline{d\rho_S \otimes \rho_E} + \overline{ \rho_S \otimes d\rho_E} + \overline{d\rho_S \otimes d\rho_E}.
\end{eqnarray}   
Using statistical properties of stochastic variables (Eqs. (\ref{eq:noise1}-\ref{eq:noise2})), we obtain 
\begin{eqnarray}
\overline{d\rho_S \otimes \rho_E} + \overline{ \rho_S \otimes d\rho_E} &=& \frac{dt}{i\hbar} [h_S + h_E , \rho_S \otimes \rho_E ], \label{eq:dd} \\
\overline{d\rho_S \otimes d\rho_E} &=& \frac{dt}{i\hbar} [ \mathbf{Q} \otimes \mathbf{B} , \rho_S \otimes \rho_E ]. \label{eq:dd2}
\end{eqnarray}
Therefore, the last term simulates the interaction Hamiltonian 
exactly and the average evolution of the total density over a time step $dt$ reads 
\begin{eqnarray}
\overline{dD} = \frac{dt}{i\hbar} [H, D], \label{eq:couple}
\end{eqnarray}
which is nothing but the exact evolution. Here, the exactness of the method is proved assuming that the density $D(t)$ 
is a single 
separable density. In practice, the total density at time $t$ is already an average over separable densities 
obtained along each stochastic path, i.e.
$\overline{D (t)} = \overline {\rho_S(t) \otimes \rho_E(t)}$. Since Eq. (\ref{eq:couple}) is valid for any density written  as $\rho_S(t) \otimes \rho_E(t)$, by summing individual contributions, we deduce that the evolution of the total density 
obtained by averaging over different paths is given 
by $i\hbar d\overline{D} = dt [H, \overline{D(t)}]$ which is valid at any time and corresponds to the exact system+environment dynamics.    
}

\subsection{Stochastic mean-field dynamics}

Here, a slightly modified version of the stochastic process is used. It incorporates part of the system/environment coupling using a "mean-field" approximation in the deterministic evolution. Following ref. \cite{Lac05}, we consider the coupled stochastic evolutions, called hereafter "Stochastic Mean-Field" (SMF) : 
\begin{widetext}
\begin{eqnarray}
\left\{
\begin{array} {lll}
d\rho_S &=& \frac{dt}{i\hbar}[h_S + \langle \mathbf{B} (t) \rangle_E \mathbf {Q} ,\rho_S] +  
d\xi_S (\mathbf{Q} - \langle \mathbf{Q}(t) \rangle_S) \rho_S + d\lambda_S 
\rho_S (\mathbf{Q} - \langle \mathbf{Q}(t) \rangle_S)
\\
\\
d\rho_E &=& \frac{dt}{i\hbar}[h_E + \langle \mathbf{Q}(t) \rangle_S \mathbf{B},\rho_E] +  
d\xi_E (\mathbf{B} - \langle \mathbf{B}(t) \rangle_E) \rho_E 
+ d\lambda_E \rho_E (\mathbf{B} - \langle \mathbf{B}(t) \rangle_E ) 
\end{array}
\right. ,
\label{eq:stocmf}
\end{eqnarray}
where 
\begin{eqnarray}
\langle \mathbf{Q}(t) \rangle_S \equiv {\rm Tr}(\mathbf{Q} \rho_S(t)), ~~~~\langle \mathbf{B}(t) \rangle_E \equiv 
{\rm Tr}(\mathbf{B} \rho_E(t)).
\end{eqnarray}
The SMF version also provides an exact reformulation of the system+environment problem. Indeed, the two terms in 
Eqs. (\ref{eq:dd}-\ref{eq:dd2}) now read
\begin{eqnarray}
\overline{d\rho_S \otimes \rho_E} + \overline{ \rho_S \otimes d\rho_E} &=& \frac{dt}{i\hbar} [h_S + \langle \mathbf{B} (t) \rangle_E  \mathbf {Q} ,\rho_S \otimes \rho_E ] + \frac{dt}{i\hbar} [h_E + \langle \mathbf{Q} (t) \rangle_S  \mathbf {B}, \rho_S \otimes \rho_E ]  \nonumber \\
\overline{d\rho_S \otimes d\rho_E} &=& \frac{dt}{i\hbar} [ (\mathbf{Q} -  \langle \mathbf{Q} (t) \rangle_S )\otimes (\mathbf{B} -  \langle \mathbf{B} (t) \rangle_E ), \rho_S \otimes \rho_E ]. \nonumber
\end{eqnarray}  
and properly recombine to recover equation (\ref{eq:couple}).   
\end{widetext}
\subsubsection{Properties of the SFM theory}
Eqs. (\ref{eq:stocmf}) have several advantages compared to the simple version (Eqs. (\ref{eq:stocmfsimple})).
First, this stochastic process automatically insures ${\rm Tr}(\rho_E) = {\rm Tr}(\rho_S) =1$ along the stochastic path.
{In addition, the inclusion of a "mean-field" term in the deterministic part
will always reduce the statistical 
dispersion compared to the simple stochastic process given by Eqs. (\ref{eq:stocmfsimple}). 
This reduction could be significant if quantum fluctuations of the coupling operators 
$\mathbf{Q}$ and $\mathbf{B}$ remain small along each path   
\cite{Lac05,Lac07}. 
Indeed, at any time, a measure of the statistical fluctuations is given by 
\begin{eqnarray}
\lambda_{stat}  &=& \overline{ {\rm Tr}\Big\{ \left(D^{\dagger} (t) - \overline{D^\dagger(t)} \right) \left(D(t) - \overline{D(t)} \right) \Big\} } \nonumber \\
&=&  \overline{ {\rm Tr}\Big\{D^{\dagger}D(t)\Big\} } - {\rm Tr} \Big\{\overline{D(t)}^2\Big\} . 
\label{eq:stat}
\end{eqnarray}
Starting from the total density associated to a pure state, 
the evolution of $\lambda_{stat}$ over a small time step is directly connected to the average quantum fluctuations of $\mathbf{Q}$
and $\mathbf{B}$, i.e.
\begin{eqnarray}
d \lambda_{stat} &=& \frac{2dt}{\hbar} 
\Big\{ \overline{(\langle  \mathbf {Q}^2 \rangle_S- \overline{ \langle  \mathbf{Q} \rangle}^2_S)} 
+ \overline{(\langle  \mathbf {B}^2 \rangle_S- \overline{ \langle  \mathbf{B} \rangle}^2_S)}  \Big\} ,
\label{eq:statmf}
\end{eqnarray}
where, we have assumed implicitly that all second moments except those given in equations (\ref{eq:noise1}-\ref{eq:noise2}) 
cancel out. For comparison, the growth of statistical fluctuations associated to the stochastic process without mean-field (Eqs. 
(\ref{eq:stocmfsimple})) reads
\begin{eqnarray}
d \lambda_{stat} &=& \frac{2dt}{\hbar} 
\Big\{ \overline{\left\langle  \mathbf {Q}^2 \right\rangle_S} +  \overline{\left\langle  \mathbf {B}^2 \right\rangle_E} \Big\}.
\label{eq:statnomf}
\end{eqnarray}
Eq. (\ref{eq:statmf}) illustrates that the number of trajectories required to simulate the system dynamics will  
depend on the importance of quantum fluctuations of coupling operators along each path. In addition, a  comparison
between Eqs. (\ref{eq:statmf}) 
and (\ref{eq:statnomf}) illustrates that the introduction of mean-field will always improve the numerical accuracy. 
}

\subsection{Reduced system evolution}
 
In Eq. (\ref{eq:stocmf}), the influence of the environment 
on the system only enters through $\langle \mathbf{B} (t) \rangle_E$. 
One expects in general to simplify the treatment by directly considering this observable 
evolution instead of the complete $\rho_E$ evolution. 
To express $\langle \mathbf{B} (t) \rangle_E$, we introduce the environment evolution operator  
\begin{eqnarray}
U_E(t,t') \equiv \exp\left( \frac{1}{i\hbar} h_E (t-t')\right).
\end{eqnarray}
Defining the 
new set of stochastic variables  $dv_E$ and $du_E$ through
$d\xi_E = dv_E - i du_E$ and $d\lambda_E = dv_E + i du_E$, the evolution of $\rho_E(t)$ can be integrated as
\begin{widetext}
\begin{eqnarray}
\rho_E(t) &=& U_E(t,t_0) \rho_E(t_0) U^\dagger_E(t,t_0)  + \int_0^t 
\frac{ds}{i\hbar}\langle \mathbf{Q} (s)  
\rangle_S~~U_E(t,s)[ \mathbf{B}  , \rho_E (s)] U^\dagger_E(t,s) \nonumber \\
&+& \int_0^t
dv_E(s)  ~~U_E(t,s) \{  \mathbf{B}  - \langle \mathbf{B}(s) \rangle_E , \rho_E(s) \}_+ U^\dagger_E(t,s) 
- i  \int_0^t du_E(s) ~~U_E(t,s)  [ \mathbf{B} - \langle \mathbf{B}(s) \rangle_E,\rho_E(s) ] U^\dagger_E(t,s) . 
\label{eq:rhoet}
\end{eqnarray} 
Introducing also the new variables $du_S$ and $dv_S$ defined as $d\xi_S = du_S -idv_S$ and 
$d\lambda_S = du_S + i dv_S$, the stochastic equation on the reduced density reads 
\begin{eqnarray} 
d \rho_S = \frac{dt}{i\hbar} \left[{\cal H}_S (t), \rho_S \right] + du_S \{ \mathbf{Q} 
-\langle \mathbf{Q} (t) \rangle_S  , \rho_S \}_+ 
- i dv_S [\mathbf{Q} -\langle \mathbf{Q} (t) \rangle_S , \rho_S]  
\label{eq:rhos}
\end{eqnarray}
with ${\cal H}_S (t) \equiv h_S + \left\langle \mathbf{B}(t) \right \rangle_E \mathbf{Q}$ and
where the source term $\left\langle \mathbf{B}(t) \right \rangle_E$ takes the exact form
\begin{eqnarray}
\left\langle \mathbf{B}(t) \right\rangle_E &=& {\rm Tr}(\mathbf{B}^I(t -t_0) \rho_E(t_0))-
\frac{1}{\hbar}\int^t_0 D(t,s) \left\langle \mathbf{Q} (s) \right\rangle_S ds 
-\int^t_0 D(t,s) du_E(s)  + \int^t_0 D_1(t,s) dv_E(s) . 
\label{eq:btime}
\end{eqnarray}
\end{widetext} 
Here, $\mathbf{B}^I(t-s) \equiv U^\dagger_E(t,s) \mathbf{B} U_E(t,s)$ while $D$ and $D_1$ are defined by:
\begin{eqnarray}
D(t,s) & \equiv & i \langle [ \mathbf{B}, \mathbf{B}^I(t-s)] \rangle_E \label{eq:d1},
\\
D_1(t,s) &\equiv& \langle  \{ \mathbf{B}  - \langle \mathbf{B}(s)\rangle_E , \mathbf{B}^I(t-s)  \}_+ \rangle_E \label{eq:d2},
\end{eqnarray}
where the environment expectation values are taken at time $s$, i.e. $\langle \cdots \rangle_S \equiv {\rm Tr} (\cdots \rho_E(s))$.
The two coupled equations, Eqs. (\ref{eq:rhos}-\ref{eq:btime}) provide an exact reformulation of the system 
evolution if $du_{S/E}$ and $dv_{S/E}$ verify
\begin{eqnarray}
\begin{array}{ccc}
\overline{du_Sdu_E} &=& \overline{dv_Sdv_E} = \frac{dt}{2\hbar},  \\
\overline{du_Sdv_E} &=& \overline{dv_Sdu_E} = 0. 
\end{array}
\label{eq:noiseusue}
\end{eqnarray}
In the following, we simply assume that the first 
term in Eq. (\ref{eq:btime}) cancels out. 
Substituting Eq. (\ref{eq:btime}) into Eq. (\ref{eq:rhos}), we finally obtain an integro-differential stochastic equation 
for the exact  system evolution where the environment effect has been incorporated through the two memory functions 
(\ref{eq:d1}-\ref{eq:d2}). 
The interesting aspect of replacing Eq.(\ref{eq:stocmf}) by 
Eqs. (\ref{eq:rhos}-\ref{eq:btime}) is that, in many physical situations, 
one can generally take advantage of 
specific commutation/anti-commutation properties of $\mathbf{B}$ as 
well as flexibility in the noise to obtain an explicit form of the memory functions. 

\section{Application to system coupled to a heat-bath}
The method is illustrated for systems coupled to an environment 
of harmonic oscillators initially at thermal equilibrium and it shows
that the present stochastic theory can be a valuable tool to simulate exactly open quantum systems.  
We take 
\begin{eqnarray}
h_E = \sum_{n}\left(\frac{p^2_n }{2m_n} + \frac{1}{2}m_n \omega^2_n x^2_n\right)
\end{eqnarray}
and $\mathbf{B} \equiv -\sum_n \kappa_n x_n$ \cite{Bre02}.
The statistical properties of stochastic variables $du_{S/E}$ and $dv_{S/E}$ specified above do not uniquely define 
the Wiener process. 
A simple prescription is to further assume 
\begin{eqnarray}
\left\{
\begin{array} {lll}
\overline{du_Sdu_S} &=& \overline{du_Edu_E} = \overline{dv_Sdv_S} = \overline{dv_E dv_E} = 0,  \\
\overline{du_Sdv_S} &=& \overline{du_Edv_E} = 0 .
\end{array}
\right.
\label{eq:statusvs}
\end{eqnarray}
There are several advantages to this choice. First, stochastic calculus are greatly simplified.
For instance, using standard techniques for system coupled to heat-bath \cite{Fey63,Cal82} 
and Ito stochastic rules,
shows that $D$ and $D_1$ depend only on the time difference $\tau = (t-s)$ and    
identify with the standard correlation functions \cite{Bre02} (see appendix \ref{app:heatbathdd1}): 
\begin{eqnarray}
\hspace*{-0.2 cm}
D(\tau) &=&  2\hbar \int^{+\infty}_0 d\omega J(\omega) \sin(\omega \tau) , \label{eq:d1exp} \\
\hspace*{-0.2 cm} D_1(\tau) &=&  2\hbar \int^{+\infty}_0 d\omega J(\omega) \coth({\hbar \omega / 2k_B T}) \cos(\omega \tau) , \label{eq:d2exp}
\end{eqnarray} 
where 
\begin{eqnarray}
J(\omega)&\equiv& \sum_n \frac{ \kappa^2_{n}} {2m_n \omega_n} \delta(\omega - \omega_n) ,
\label{eq:spectral}
\end{eqnarray}
denotes the spectral density.
No approximation are made to obtain above equations, therefore the average over different stochastic paths
match the exact evolution of the system, including 
all non-Markovian effects.

\subsection{Equivalent Stochastic Schroedinger Equation formulation}

Several works, based on the influence functional method \cite{Sto02,Sha04,Yan04}
have led to similar stochastic equations for the reduced density. 
For instance, authors of ref. \cite{Sha04,Yan04} use an evolution 
of $\left\langle  \mathbf{B}(t) \right\rangle_E$ where the second term in Eq. (\ref{eq:btime}) is absent. 
As demonstrated below, this term is of crucial importance for applications.
In ref. \cite{Sto02} and in refs. \cite{Bre04a,Bre04b,Lac05} a stochastic formulation of the exact system+environment 
is given in terms of the Stochastic Schroedinger Equation (SSE) technique.  
Thanks to the additional stochastic rules (\ref{eq:statusvs}), Eq. (\ref{eq:rhos}) has also its SSE
counterpart, where system densities are replaced by 
$\rho_S = \left| \phi_1 \right\rangle \left\langle \phi_2 \right|$ and wave functions 
evolve according to 
\begin{eqnarray}
\left\{
\begin{array} {ccc}
d \left| \phi_1 \right\rangle &=&  \{\frac{dt}{i\hbar}{\cal H}_S (t)  + d\xi_S 
(\mathbf{Q} - \langle \mathbf{Q} (t) \rangle_S ) \} \left| \phi_1 \right\rangle \\
\\
d\left\langle \phi_2 \right| &=& \left\langle \phi_2 \right|  \{-\frac{dt}{i\hbar}{\cal H}_S (t)  + d\lambda_S 
(\mathbf{Q} - \langle \mathbf{Q} (t) \rangle_S ) \} 
\end{array}
\right.
.
\end{eqnarray}
where the bath effect is again incorporated through the mean-field kernel.  

\subsection{Application to the spin-boson model}

We illustrate the proposed technique to the spin-boson model. 
This model can be regarded as one of the simplest quantum open system coupled to a heat bath 
\cite{Leg87} which could not be integrated exactly. In addition, it has been often used as a benchmark for 
theories of open quantum systems \cite{Leg87,Egg94,Dio98,Wei99,Bre01,Sto02,Sha04}.   
The system  and coupling Hamiltonians are respectively chosen as 
\begin{eqnarray}
h_S = \hbar \omega_0 \sigma_x + 
\hbar \varepsilon  \sigma_z, 
~~~~ 
h_I =  \sigma_x \otimes \mathbf{B} \nonumber,
\end{eqnarray}
where the $\{ \sigma_i \}_{i=x,y,z}$ are the standard Pauli matrices. 
In the spin-boson model, the numerical solution of equation (\ref{eq:rhos}) for the system density, 
is equivalent to solving three non-linear coupled equations for the $\left\langle \sigma_{i} \right\rangle_S$.
\begin{figure} [htbp]
\epsfig{file=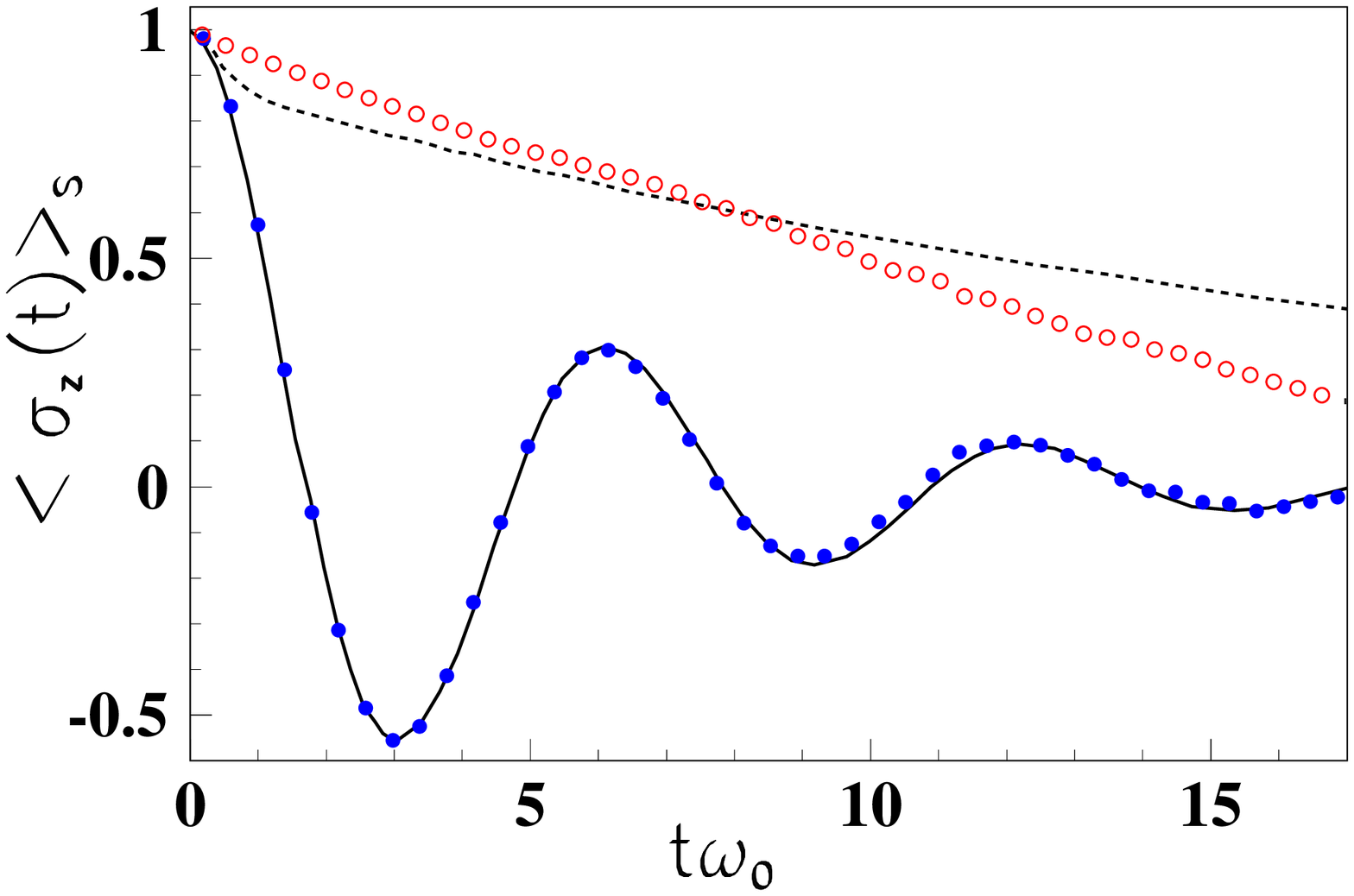,width=\linewidth,clip}
\caption{Evolution of $\left\langle \sigma_z(t) \right\rangle_S$ (assuming 
$\left\langle \sigma_z(0) \right\rangle_S = 1$) as a function of time obtained through the average
of paths simulated with Eq. (\ref{eq:rhos}) with the Markovian process described by Eqs. (\ref{eq:noiseusue}-\ref{eq:statusvs}) 
and memory functions given by Eq. 
(\ref{eq:d1exp}-\ref{eq:d2exp}). 
We assume that $\displaystyle J(\omega) = \eta 
\omega \frac{\Delta^2_c}
{\Delta^2_c + \omega^2}$. Two sets of parameters are used. In both cases, 
 $\Delta_c = 5 \omega_0$, $ k_B T = 2 \hbar\omega_0$ and $\hbar \varepsilon=0$. The filled and open circles correspond
respectively to $\pi \eta = 0.2 \hbar\omega_0$ and $\pi \eta = 4 \hbar\omega_0$
and are compared respectively with the solid and dashed line obtained with the same set of
parameters in Fig. 2 of ref. \cite{Yan04}. Results are obtained with $2 \times 10^4$ trajectories.}
\label{fig1:bath}
\end{figure}
Fig. \ref{fig1:bath} shows examples of dynamical evolution of $\overline{\langle \sigma_z(t) \rangle_S}$ 
obtained using Eq. (\ref{eq:rhos}) and averaging over stochastic 
trajectories both for weak (filled circle)
and strong (open circle) coupling. 
Results are compared with the Hierarchical approach proposed in ref. 
\cite{Yan04}. This deterministic approach provides an alternative {\it a priori} exact formulation of open quantum 
systems dynamics and was originally motivated by numerical difficulties encountered in the stochastic theory 
proposed in ref. \cite{Sha04}. Such difficulties do not occur 
in the present simulation and much less stochastic trajectories seem to be needed to accurately describe 
the dynamical evolution. Only $2 \times 10^4$ trajectories have been used to obtained Fig. \ref{fig1:bath} leading to
statistical errors close to zero (for comparison see discussion in \cite{Zho05}). 
The computer time for the two figures was less than an hour for the weak 
coupling case up to several hours for the strong coupling case on a standard personal computer. The difference in 
computing time comes from the fact that smaller numerical time step should be used as the coupling strength 
increases to achieve good numerical accuracy,
the main difficulty being to properly evaluate time integrals in Eq. (\ref{eq:btime}).  
Denoting the time step by $\Delta t$, $\Delta t \omega_0 = 1.2 \times 10^{-3}$ and 
$\Delta t \omega_0 = 2.2 \times 10^{-4}$ have been used for weak and strong coupling respectively. 
 
In the weak coupling case, results of our stochastic scheme displayed in Fig. \ref{fig1:bath} (filled circles) 
perfectly match the result of ref. \cite{Yan04} (solid line). In contrary to ref. \cite{Zho05}, statistical errors 
remain small even for long time evolution. The difference in numerical accuracy can be assigned to the 
second term in Eq. (\ref{eq:btime}) which turns out to be crucial for numerical implementation. Stochastic 
simulations for strong coupling parameters (open circles) slightly differ from the results obtained with 
the hierarchical approach in ref. \cite{Yan04}. 
The numerical convergence of the stochastic simulation presented in Fig. \ref{fig1:bath} has been 
checked. Therefore, the difference might be due to the fact that the numerical accuracy depends on the truncation scheme used in the 
hierarchy, even though the 
method of ref. \cite{Yan04} is exact. 

\subsection{Comparison with the Time-convolutionless method (TCL)}

\begin{figure} [htbp]
\epsfig{file=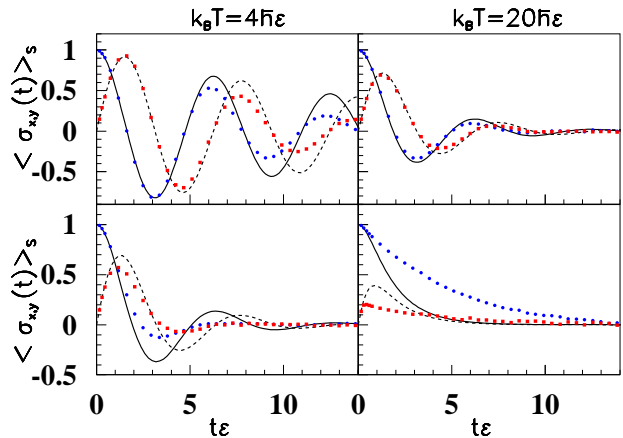,width=\linewidth}
\caption{Average evolution of $\left\langle \sigma_x(t) \right\rangle_S$ (filled circles) and 
$\left\langle \sigma_y(t) \right\rangle_S$ (filled squares) as a function 
of time.
The initial condition correspond to $\left\langle \sigma_x(0) \right\rangle_S = 1$. In all cases, $\hbar\omega_0 = 0$
and $\Delta_c = 10 \varepsilon$. $k_B T = 4 \hbar\varepsilon$ and $k_B T = 20 \hbar \varepsilon$  are used 
respectively for calculations presented in the left and right column. In both cases, upper panels present results 
for weak coupling ($\pi \eta = 0.2 \hbar\varepsilon$) while in lower panels a stronger coupling is considered
($\pi \eta = 1.0 \hbar\varepsilon$). Simulations have been performed with $4 \times 10^4$ trajectories. For all cases, 
dynamical evolutions of the $x$
and $y$ spin components obtained with the TCL2 method \cite{Bre99,Bre02} are displayed by 
solid and dashed lines respectively.}
\label{fig2:bath}
\end{figure}

The possibility to simulate exactly the system dynamics can also serve as a benchmark for other 
methods. For instance, we compared the exact stochastic scheme with the approximate 
Time-Convolutionless (TCL) projection operator method of ref. \cite{Bre99,Bre02}. 
Figure \ref{fig2:bath} presents the results of the exact stochastic simulation compared with the 
TCL2 method applied to the spin-boson model in ref. \cite{Bre99}. 
In this figure different cases corresponding to either low or high temperature regime and weak 
or strong coupling are presented. We see that the best agreement is obtained in the weak coupling and 
high temperature case. In general, the TCL2 method compares well with the exact simulation 
if the coupling is rather small. As the coupling increases (lower panels of figure \ref{fig2:bath}), 
the difference between the TCL technique and the exact method increases. Note that the TCL method seems 
to systematically underestimate the damping. Note finally that the use of TCL4 
equations instead of TCL2 does not improve the comparison.

\subsection{Discussion of approximate system evolution obtained with real noise and Hermitian system densities 
in stochastic evolution}

Applications presented above use specific constraints on the Markovian process given by 
Eqs. (\ref{eq:statusvs}). This prescription greatly simplifies stochastic calculus. For instance, simple 
exact expressions have been obtained for $D(t,s)$ and $D_1(t,s)$ when the system is coupled to a heat-bath
of harmonic oscillators (Eqs. (\ref{eq:d1exp}-\ref{eq:d2exp})). The main consequence of Eqs. (\ref{eq:statusvs})
is that $du_{S/E}$ and $dv_{S/E}$ should be {\it complex} stochastic variables leading to non Hermitian densities 
along paths. As illustrated above, such a stochastic process could be used to simulate exactly 
the reduced density evolution. The main disadvantage of non-Hermitian densities is however that 
system observables could hardly be interpreted. We discuss here the possibility to perform stochastic 
evolution of reduced Hermitian densities.  

\begin{figure} [htbp]
\epsfig{file=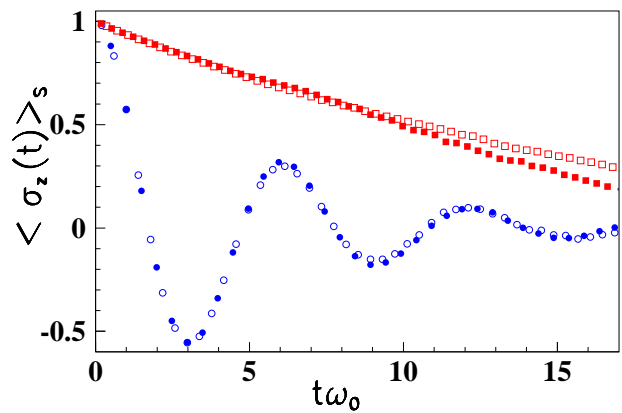,width=\linewidth}
\caption{Results obtained in the weak and strong coupling case with the exact 
stochastic simulation using complex noises and non-hermitian (filled circles) are compared with the approximate simulation 
(open circles) using real noise and hermitian densities along each paths. In this figure, the values of parameters are the same 
as in Fig. \ref{fig1:bath}. }
\label{fig3:bath}
\end{figure}

\begin{figure} [htbp]
\epsfig{file=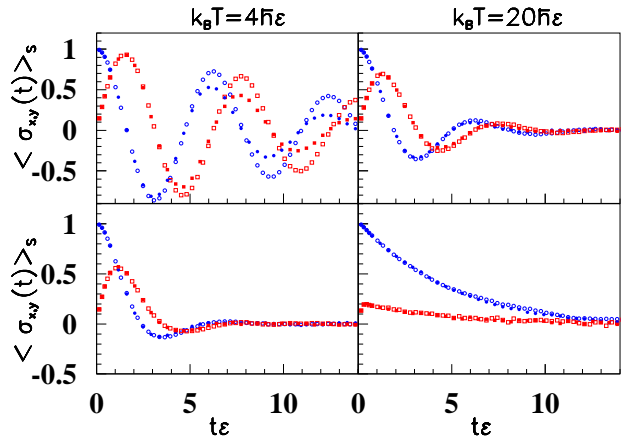,width=\linewidth}
\caption{Average evolution of $\left\langle \sigma_x(t) \right\rangle_S$ (filled circles) and 
$\left\langle \sigma_y(t) \right\rangle_S$ (filled squares) as a function 
of time obtained with the exact stochastic scheme using the same sets of parameters as in figure 
\ref{fig2:bath}. In each cases, dynamical evolutions of the $x$
and $y$ spin components obtained with the approximate stochastic simulation using real noise are 
respectively displayed with open circles and open squares.}
\label{fig4:bath}
\end{figure}

Relaxing the constraints given by Eqs. (\ref{eq:statusvs}), authorizes to choose $du_{S/E}$ and $dv_{S/E}$ as 
real stochastic variables, which automatically insures that $\rho_S(t)$ and $\rho_E(t)$ remain hermitian. 
This alternative has however two major drawbacks. First, one cannot anymore have an equivalent SSE picture.
Second, while $D(t,s)$ still identifies with Eq. (\ref{eq:d1exp}), no simple exact expression can be worked out 
for $D_1(t,s)$. However, since this kernel is a functional of $\rho_E(s)$, a hierarchy of 
more and more accurate approximations could be obtained by successive replacements of $\rho_E(s)$ into Eq. (\ref{eq:d1})
by its integral expression, Eq. (\ref{eq:rhoet}). 
In the present work, we concentrate on the simplest case where
$\rho_E(s)$ is replaced by $\rho_E(s) \simeq U_E(t,t_0) \rho_E(t_0) U^\dagger_E(t,t_0)$  in the time integral of the memory kernel.
In this limit, $D_1(s,t)$ also reduces 
to Eq. (\ref{eq:d2exp}). 
By doing this approximation, the stochastic process is not exact 
anymore. 
Fig. \ref{fig3:bath} presents 
a comparison of the exact stochastic simulation obtain with complex noise (filled circles) 
and the approximate case with real noise (open circles). The parameters of the spin-boson model 
are the same as in Fig. \ref{fig1:bath}.
This figure shows that the approximate scheme with real noise is very close 
to the exact simulation even in the strong coupling limit. In the latter case, only at very large time, the two simulation 
start to deviate slightly from one another. For completeness, approximate stochastic simulations
obtained for cases presented in figure \ref{fig2:bath} are compared to the exact scheme in figure \ref{fig4:bath}. 
We see that except for the weak coupling and low temperature case, the approximate simulation is very close to the exact 
case. 
It is worth mentioning that approximate simulation presented here uses the simplest prescription 
for $D_1(s,t)$. Therefore, improved description could {\it a priori} be obtained using better approximation of 
$D_1(s,t)$ obtained with the method described above.    
This example is very encouraging and provides a new method to simulate open systems with a stochastic 
process preserving the hermitian properties of the system density.  

\section{Conclusion}

The results obtained with the stochastic theory for the spin-boson model are very encouraging.
The theory turns out to be accurate not only to simulate short but also long-time dynamics and does not seem 
to suffer from numerical instability quoted in ref \cite{Zho05}. 
It is worth mentioning that optimization techniques proposed in ref. \cite{Lac05} which are not used here, 
can be implemented to further reduce the number of stochastic paths. 
Besides the numerical aspects, the link with classical dissipations dynamics 
could be easily made, similarly to ref. \cite{Sto02}.

Our stochastic approach could be generalized to any initial conditions that could be 
written as a mixing of separable densities, i.e. $D(t_0) = \sum_n W_n D^{(n)}(t_0)$ with $D^{n} = \rho^{n}_S \otimes \rho^{n}_E$ 
where the $W_n$  are complex coefficients. 
Then, the complete exact dynamics is recovered by both averaging over trajectories starting from each $D^{(n)}(t=0)$ individually 
and averaging over the initial conditions. The theory is also not restricted to real time
evolution. Statistical properties of the system+environment could be studied by 
considering imaginary time propagation, i.e. $i dt/\hbar \rightarrow \beta$. Imaginary time propagations lead naturally to
densities written as a mixing of separable densities and could then serve as initial conditions for real time evolution.  
By combining both imaginary time and real time 
propagation, general physical problems similar to those depicted in ref. \cite{Sto02} can be addressed. 
The main limitation of the technique is clearly the choice of coupling operator $\mathbf{B}$ which should 
give simple memory functions (Eq. (\ref{eq:d1} - \ref{eq:d2})). 
It is however worth mentioning that most of 
the coupling operators used in the context of open quantum systems already enter into 
this category \cite{Car93,Wei99,Gar00,Bre02}. 
We believe that the stochastic theory presented in this paper can be a valuable tool to access 
exactly the dynamics of more complex open quantum systems. We presented here specific 
applications on systems coupled to a heat-bath.  
The approach can however be applied to various 
types of environments and couplings which might be of great interest to 
address dissipation, measurement and/or decoherence problems in quantum systems.

\acknowledgments{The author thank G. Adamian,  N. Antonenko, S. Ayik, and B. Yilmaz
for valuable discussions and C. Simenel, G. Hupin for his careful reading of  the manuscript. }

\appendix 

\section{Proof of Eq. (\ref{eq:btime}) for a heat-bath of Harmonic oscillators}
\label{app:heatbathdd1}

We give here a proof of Eq. (\ref{eq:btime}) where $D$ and $D_1$ identify 
with Eqs. (\ref{eq:d1exp}-\ref{eq:d2exp}). The environment is assumed to be
a set of harmonic oscillators, labeled by "$n$" associated to creation/annihilation 
($a^\dagger_n$,$a_n$), i.e.
\begin{eqnarray}
H =\sum_n \hbar \omega_n \left(a^\dagger_n a_n +\frac{1}{2} \right).
\end{eqnarray} 
If thermal equilibrium is initially assumed, the environment density writes as a product 
of densities of each oscillator, $\rho_E = \Pi_n \rho_n$ where each density can be written
as Gaussian operators (see for instance \cite{Gar00}) determined by first and second 
moments of the $(a^\dagger_n,a_n)$. 
The time evolution of the environment is given by Eq. 
(\ref{eq:stocmf}) where $\mathbf{B} \equiv -\sum_n \kappa_n x_n$ and where the fluctuating variables 
verify (according to Eq. (\ref{eq:statusvs})):
\begin{eqnarray}
\overline{d\lambda_E d\lambda_E} = 
\overline{d\xi_E d\xi_E} = 
\overline{d\lambda_E  d\xi_E} = 0. 
\end{eqnarray} 
Above prescription and the specific form of $\mathbf{B}$ induces important simplifications listed below. First, 
the initial product form of the environment density is preserved along the stochastic paths where 
each oscillator density verifies ${\rm Tr}(\rho_n) = 1$ and where for each pairs of densities 
$d(\rho_n \rho_m) = \rho_n d\rho_m + d\rho_n \rho_m$. Second, due to the linear coupling operator, 
the Gaussian nature of initial densities is also preserved along paths. Therefore, we can equivalently 
solve the density equation of motion or follow first and second moments of each $(a^\dagger_n,a_n)$ in time. 
Here we consider the second strategy.

From the $\rho_E$ evolution, the equation of motion of each pairs $\langle a^\dagger_n \rangle = {\rm Tr}(\rho_E a^\dagger_n)$ and 
$\langle a_n \rangle = {\rm Tr}(\rho_E a_n)$ reads:
\begin{widetext}
\begin{eqnarray}
d\left\langle a_n \right\rangle &=& -i\omega_n dt \left\langle a_n \right\rangle + 
\frac{dt}{i\hbar} c_n \left\langle  {\mathbf Q} \right\rangle 
+ c_n  \left\{ d\xi_E 
\left( \sigma^{(n)}_{+-}(t) + \sigma^{(n)}_{--}(t) + 1 \right) + d\lambda_E 
\left( \sigma^{(n)}_{+-}(t) + \sigma^{(n)}_{--}(t)  \right) \right\} , \label{eq:an} \\
d\left\langle a^\dagger_n \right\rangle &=& +i\omega_n dt \left\langle a^\dagger_n \right\rangle - 
\frac{dt}{i\hbar} c_n \left\langle  {\mathbf Q} \right\rangle + c_n  \left\{ d\xi_E 
\left( \sigma^{(n)}_{+-}(t) + \sigma^{(n)}_{++}(t) \right) + d\lambda_E 
\left(  \sigma^{(n)}_{+-}(t) + \sigma^{(n)}_{++}(t) + 1\right) \right\} , \label{eq:adn} \nonumber
\end{eqnarray}  
\end{widetext}
where we have introduced the notation $c_n \equiv -\kappa_n / \sqrt{2\eta_n}$ and $\eta_n = m_n \omega_n/\hbar$. Here, $\sigma^{(n)}_{++}$, 
$\sigma^{(n)}_{--}$ and $\sigma^{(n)}_{+-}$ denote the second moments of the $a^\dagger_n$, $a_n$ operators:  
\begin{eqnarray}
\sigma_{+-}^{(n)} &\equiv& \left\langle a^\dagger_n a_n \right\rangle - \left\langle a_n \right\rangle\left\langle a^\dagger_n \right\rangle = \sigma_{-+}^{(n)} -1  , \nonumber \\
\sigma_{--}^{(n)} &\equiv & \left\langle a_n a_n \right\rangle - \left\langle a_n \right\rangle\left\langle a_n \right\rangle , \nonumber \\
\sigma_{++}^{(n)} &\equiv & \left\langle a^\dagger_n a^\dagger_n \right\rangle - \left\langle a^\dagger_n \right\rangle\left\langle a^\dagger_n \right\rangle . \nonumber 
\end{eqnarray}
According to the stochastic environment dynamics, we can show that these moments  simply evolve as 
\begin{eqnarray}
\left\{
\begin{array} {lll}
d\sigma^{(n)}_{--} &=& -2i\omega_n dt \sigma^{(n)}_{--} , \\
d\sigma^{(n)}_{++} &=& +2i\omega_n dt \sigma^{(n)}_{++} , \\ 
d\sigma^{(n)}_{+-} &=& 0 .
\end{array}
\right.
\end{eqnarray} 
Since we assume that each oscillator is initially at thermal equilibrium, we deduce that second moments 
are constant in time with $\sigma^{(n)}_{--}(t) = \sigma^{(n)}_{++}(t) =0$, while 
$\sigma^{(n)}_{-+}(t) = \sigma^{(n)}_{+-} + 1 = (\bar N(\omega_n)+ 1)$. Here, we have introduced the 
standard function \cite{Gar00}: 
$(2  \bar N (\omega_n) + 1) = \coth\left(\hbar \omega_n /(2k_B T) \right)$. 
Substituting in Eqs. (\ref{eq:an}-\ref{eq:adn}) and using standard integration techniques \cite{Bre02} leads 
finally to   
\begin{widetext}
\begin{eqnarray}
\left\langle a_n(t) \right\rangle = e^{-i\omega_n t}\left\langle a_n(0) \right\rangle +\frac{c_n}{i\hbar} 
\int_0^t  e^{-i\omega_n (t-s)} \left\langle  {\mathbf Q}(s) \right\rangle ds +  
c_n \int_0^t e^{-i\omega_n (t-s)}\left\{ d\xi_E(s)  \left( \bar N (\omega_n) + 1 \right) + d\lambda_E(s)  \bar N (\omega_n)  \right\} ,
\nonumber \\
\left\langle a^\dagger_n(t) \right\rangle = e^{+i\omega_n t}\left\langle a^\dagger_n(0) \right\rangle -\frac{c_n}{i\hbar} 
\int_0^t  e^{+i\omega_n (t-s)} \left\langle  {\mathbf Q}(s) \right\rangle ds + 
c_n \int_0^t  e^{+i\omega_n (t-s)}\left\{ d\xi_E(s)   \bar N (\omega_n) + d\lambda_E(s) 
 (\bar N (\omega_n)+ 1 )  \right\} .
\end{eqnarray}
Accordingly, each position operator entering in  $\left\langle  {\mathbf B}(t) \right\rangle$ reads: 
\begin{eqnarray}
\left\langle x_n(t) \right\rangle 
&=& \frac{1}{\sqrt{2\eta_n}} \Big\{ \left\langle a_n(0) \right\rangle e^{-i\omega_n t} + 
\left\langle a^\dagger_n(0) \right\rangle e^{+i\omega_n t} 
\Big\} 
+  \frac{\kappa_n}{\hbar \eta_n} \int^t_0 \sin(\omega_n [t-s])  \left\langle {\mathbf Q} (s) \right\rangle ds , \nonumber \\
&& - \frac{\kappa_n}{2\eta_n} \int^t_0 \left\{
\cos(\omega_n [t-s]) (2  \bar N (\omega_n) + 1) \left[   d\xi_E(s) + d\lambda_E(s) \right] -i
\sin(\omega_n [t-s]) \left[   d\xi_E(s) - d\lambda_E(s) \right]
\right\} . \nonumber
\end{eqnarray}
Assuming the initial conditions $\left\langle a_n(0) \right\rangle = \left\langle a^\dagger_n(0) \right\rangle =0$, substituting in the 
expression of $\left\langle  {\mathbf B}(t) \right\rangle = -\sum_n \kappa_n \left\langle x_n(t) \right\rangle$ and 
introducing the spectral density (Eq. (\ref{eq:spectral})), we finally recover equation (\ref{eq:btime}) where $D$ and $D_1$ are respectively
given by Eqs. (\ref{eq:d1exp}) and (\ref{eq:d2exp}).
\end{widetext}
\bibliography{paper_heatbath_ver4.bbl}
\end{document}